\documentclass[11pt]{article}
\usepackage{amssymb}
\begin{document}

\centerline{\Large\bf Scattering from small colloidal particles}
\centerline{\Large\bf in a semidilute polymer solution}
\vspace{0.5in}

\centerline{\large\bf Richard P. Sear\footnote{\large
Present address: Department of Chemistry and Biochemistry,
University of California Los Angeles, Los Angeles, CA 90024, USA.
email: sear@chem.ucla.edu}}

\vspace{1.0in}
\centerline{\large FOM Institute for Atomic and Molecular Physics,}
\centerline{\large Kruislaan 407, 1098 SJ Amsterdam, The Netherlands}
\vspace{0.5in}

%\centerline{\today}
%\vspace*{0.8in}
%{\large PACS:~~~61.25.Hq, 83.70.Hq}

%{\large Running title: Particles in semidilute polymer}

\begin{abstract}

The correlations between the segments of
a semidilute polymer solution are found
to induce correlations in the positions of small particles added to
the solution. Small means a diameter much less than the polymer's
correlation length.
In the presence of polymer the particles
behave as if they attracted each other.
It is shown how the
polymer's correlation length may be determined from a scattering
experiment performed on the spheres.

\end{abstract}

%\newpage

\section{Introduction}

Small colloidal particles in a semidilute polymer solution are
shown to interact via a long ranged, attractive potential of mean
force. This potential of mean force \cite{hill60} is an effective
interaction potential between the particles due to the presence of the
polymer.
The particles are small with respect to the polymer's correlation
length and, as it is this length which determines the range of the
attractive potential of mean force, the attractive interactions are
long ranged with respect to the direct interaction between
the spheres.
The effective attractions
between the colloidal particles are a result of the
correlations between the
segments in a semidilute polymer solution \cite{degennes}.
The segments of a polymer molecule in a good solvent follow a self-avoiding
walk.
Because of this the fraction
of the volume of the solution occupied by the polymer segments
is highly non-random: if there is a segment at one point then
the average density of polymer nearby is much higher than the average
density of segments. This implies that the volume {\it not} occupied by
the polymer segments is also highly correlated; if
you imagine a picture in which every point is either black or
white, then if there is a pattern to the white areas then there must
be a pattern to the black areas. But the particles are only
free to move in this volume which is free of polymer; they
are excluded from the volume occupied by the polymer. So the
particles move in a fraction of the solution's volume which has long
range correlations and so their positions have long range correlations.
In fact the correlations between the spheres
resemble those produced by a long ranged
attraction.
As the correlations in the spheres reflect those in the polymer,
scattering experiments \cite{cannell87}
on the colloidal particles may be used to probe the structure of a
semidilute polymer solution; in particular they provide a way of
measuring the correlation length of the polymer.

The colloidal particles could be
protein molecules \cite{abbott91,abbott92}, surfactant micelles
\cite{clegg94,robb95}, or
synthetic, polymer or silica, spheres
\cite{pusey91,lekkerkerker95}.
Both protein molecules and micelles typically have
diameters of a few nm. The direct interaction between particles is taken
to be purely repulsive, there are no attractive interactions.
The polymer is considered to be in a good
solvent and is therefore swollen
due to self-interactions \cite{degennes}. The polymer's radius
of gyration is much larger than the diameter of the particles,
$\sim 10$nm or more. 
See Refs.
\cite{vrij76,yethiraj92,meijer94} for examples of work where
the radius of gyration is comparable to or smaller than
than the diameter of the particles.
Recent work of Khalatur {\it et al.} \cite{khalatur97} has dealt
with similar mixtures to those considered here using RISM integral
equations. It is more numerical and less intuitive than here. The RISM
approach is also best suited to higher polymer densities than those
considered here,
say volume fractions of polymer of 10\% and more.
Our simple scaling approach breaks down at these densities.
All interactions are assumed to be excluded volume interactions, that is
two polymer segments may not occupy the same volume and likewise for
two particles or a particle and a polymer segment. This is
reasonable if the solvent is good for the polymer,
and the polymer does not absorb onto the particle.

All three interactions are excluded volume and therefore
there are no energy scales, apart from the temperature $T$.
Therefore, the behaviour is a function only of the length scales
of the mixture. The length scale of the particles is just their
diameter $D$.
A pure semidilute polymer solution has only one
relevant length scale, the correlation length $\xi$ \cite{degennes},
which is roughly the distance between interactions between segments
on different polymer chains.
The radius of gyration is not a
relevant length scale as it is much larger than $\xi$ and
the chain loses its correlations and `forgets' which polymer it belongs to
over a distance of $\xi$.
The mixture then has
only two relevant length
scales: $D$ and $\xi$ \cite{degennes79,odijk96,sear97}.
This means that the particle--polymer interaction
depends only on their
ratio: $D/\xi$.

\section{Theory}

The polymer molecules consist of flexible linear chains of segments
of size $a$. These segments are much smaller than the spherical
particles, $D\gg a$ and so the volumes from which the segments exclude
spheres overlap. Each segment excludes a sphere from a volume
$\sim(D+a)^3$ centred on itself but as the segments are part of a
chain there are always two segments $a$ away from any segments,
another two $\sim 2a$ away, etc.. 
This problem of the excluded volumes overlapping can be reduced 
if the polymer segment length is rescaled from $a$ to $D$. 
This is illustrated in Fig. 1, which
shows a sphere and a part of a polymer chain.
As semidilute polymer solutions are
scale invariant \cite{degennes} at length scales much less than
the correlation length $\xi$ we are free to perform this rescaling.
Now that the polymer segment length is $D$ the excluded volumes
of adjacent segments overlap much less strongly, although they
still overlap. Each rescaled segment consists
of many segments, for $D\gg a$, and so acts like a polymer coil of the same
size as the particles. The particles cannot penetrate coils
of their own size \cite{joanny79} and so the segments exclude
particles from a volume $\sim D^3$. The ratio of the number of segments of
size $D$ to those of size $a$ is $\sim(a/D)^{5/3}$ if the polymer
is in a good solvent.

From now on we only consider the scaled polymer segments of size $D$,
see Fig. 1.
Treating the polymer as a sequence of rescaled segments of size $D$
makes it clear that the only relevant length scales are $D$ and the polymer's
correlation length.
So, the polymer density $c_P$ is the density of these polymer
segments. These segments exclude each other from a volume $\sim D^3$
just as dilute polymer coils exclude each other
from a volume of order of their radius of gyration cubed.
We have approximated
the interaction between a segment and a sphere by a hard-sphere
interaction of diameter $D$; the same interaction as between the spheres.
This is an approximation of course;
the segments do not interact with a completely hard potential, they are
only very approximately spherical and they are connected into chains.

\subsection{Scaling theory for semidilute polymers}

First, we briefly summarise the scaling theory for polymer solutions in
good solvents, it is described in detail by de Gennes \cite{degennes}.
Our spheres will be added to a semidilute solution of polymer
molecules in a good solvent. As noted above we have rescaled the polymer
segment length from $a$ to $D$.
The size of an single isolated polymer coil of $n_D$ segments
is measured by its radius of gyration $R_G$ which is given by
$R_G\sim Dn_D^{3/5}$, in a good solvent  \cite{degennes}.
The correlations (fluctuations) in
an isolated polymer coil extend across the entire coil; the correlation
length $\xi$ is thus $\sim R_G$.
When polymer coils
overlap, interactions between different chains reduce the
correlation length $\xi$. The overlap
volume fraction $c^*$ is
$c_P^*=D^{-3}n_D^{-4/5}$.
This volume fraction is the boundary between the dilute and semidilute
regimes; below $c^*$ the interactions between chains can be treated
as a perturbation but above $c_P^*$ the interaction between a pair of chains
is $\gg k_BT$. In the dilute regime $\xi=R_G$, and
in the semidilute regime.
$\xi=D^{-5/4}c_P^{-3/4}$.

We require the pair distribution function for the polymer
segments $g_P(r)$, where $r$ is the magnitude of the separation of the
two segments \cite{hansen86}. It is,
\begin{equation}
g_P(r)\sim\left\{
\begin{array}{ll}
0 & ~~~~~~ r\lesssim D \\
\left(\frac{\xi}{r}\right)^{4/3} & D\lesssim r\lesssim\xi \\
1+\frac{\xi\exp(-r/\xi)}{r} & ~~~~~~ \xi\lesssim r
\end{array}\right. .
\label{gofrp}
\end{equation}
Note that we define our pair distribution function differently from
Ref. \cite{degennes}. The first line of Eq. (\ref{gofrp}) just reflects the
mutual impenetrability of polymer segments, the second is the density of
a self-avoiding random walk, and the
third line reflects the screening of correlations
at distances greater than $\xi$.
Note that for $r\ll\xi$ the dominant
contribution to $g_P$ is the contribution from a single chain.

\subsection{Scaling theory for spheres in a semidilute polymer solution}

In order to examine the pair correlations and hence the scattering
from spheres in a semidilute solution we require the pair
distribution function of the spheres, $g_S(r)$.
For $r<D$ this is 0 due to the hard-sphere interaction but
for $r>D$ $g_S$ is not equal to 1 due to the fact that the cavities
in the polymer solution are correlated; see \cite{reiss92}.
The cavities are the points in the polymer solution where there
is no polymer within a distance $D$, and so where it is possible to insert
a sphere \cite{reiss92}.
In the limit of very low sphere density $c_S$
we need only consider the two spheres which
form the pair; there are no other spheres present in this limit and
so the correlations between the spheres are entirely due
to correlations in the cavities:
\begin{equation}
\lim_{c_S\rightarrow0}g_S(r)=
\left\{
\begin{array}{ll}
0 & ~~~~~~ r<D\\
y_P(r)&  ~~~~~~ r>D
\end{array}\right. .
\label{gy}
\end{equation}
where $y_P(r)$
is the cavity--cavity correlation function \cite{hansen86},
equal to the probability of
inserting a pair of spheres a distance $r$ apart divided by the square
of the probability of inserting one sphere.
The probability of inserting a single sphere will be denoted by $\alpha$.
Below we start by estimating $\alpha$ and then go on to consider
the probability of being able to insert a pair of spheres which yields an
estimate of $y_P$.

The concentration of segments is $c_P$
and each segment excludes spheres from a volume of
$\sim D^3$. If we ignore the fact the segments are connected
then as the polymer density is low,
we can also neglect the overlaps of excluded volumes of nonadjacent
segments
and we arrive at the conclusion that the fraction of
the volume denied to the spheres is $c_PD^3$ \cite{degennes79,odijk96}. Using
$\xi=D^{-5/4}c_P^{-3/4}$ this fraction can also be written as
$(D/\xi)^{4/3}$, and the fraction of the volume available to
the spheres, $\alpha$, is
\begin{equation}
\alpha \sim 1 - \left(\frac{D}{\xi}\right)^{4/3}.
\label{alpha}
\end{equation}
Equation (\ref{alpha}) has a probabilistic interpretation which will be
used below. It is the probability that there is no polymer segment
within $D$ of a randomly chosen point in the solution.

In order to calculate the pair distribution function between spheres, $g_S$,
we relate it, using Eq. (\ref{gy}), to the distribution function
between pairs of cavities $y_P$, i.e., pairs of points with
no polymer segments  within a distance $D$ of either of them.
Then we express $y_P$ in terms of $\alpha$ and $g_P$, both of
which are known, Eqs. (\ref{alpha}) and (\ref{gofrp}).
In order to relate $y_P$ to $g_P$ we use the fact that for a pair of
points, labeled 1 and 2,
there are 4 possibilities: 1 and 2 both have polymer segments
within a distance $D$; 1 does but 2 does not; 2 does but 1 does not; and
neither does.
Therefore, the probability of the fourth possibility occurring, which is
$\alpha^2y_P$, is equal to 1 minus the sum of the probabilities of the
other possibilities.
The probability of both points being within $D$ of a polymer segment
is the square of the probability of one point having a nearby segment,
$(D/\xi)^{8/3}$, times $g_P(r)$, where $r$ is the magnitude of
the separation ${\bf r}$ between points 1 and 2.
In deriving this result we have integrated over spheres of
diameter $D$ around both points; therefore, it is only valid
for $r>2D$ but we will use it for $r>D$.
This is a very minor approximation, $y_P$ still has the correct
behaviour near $r=D$.
The probability of point 1 being near a segment and point 2 not being near
a segment, is
the probability of point 1 being near a segment, $(D/\xi)^{4/3}$,
times the probability of there being no polymer segment near
a point ${\bf r}$ away from a segment $1-(D/\xi)^{4/3}g_P(r)$. This
second probability is just 1 minus the probability of there being
a segment at a separation ${\bf r}$ from another segment.
So,
\begin{equation}
\left(1-(D/\xi)^{4/3}\right)^2y_P(r)\sim
1-(D/\xi)^{8/3}g_P(r)-2(D/\xi)^{4/3}\left(
1-(D/\xi)^{4/3}g_P(r)\right) ~~~~~~ r>D,
\label{y1}
\end{equation}
and
\begin{equation}
y_P(r)\sim1+\frac{(D/\xi)^{8/3}}{\left(1-(D/\xi)^{4/3}\right)^2}
\left(g_P(r)-1\right) ~~~~~~ r>D.
\label{y}
\end{equation}
Note that $\alpha^2y_P(D)\sim\alpha$ which is at it should be, the probability
of inserting two particles next to each other is almost the same as the
probability of being able to insert one particle.
The probability of inserting a sphere close to where a sphere has
already been successfully inserted is higher than at a randomly
chosen point in the polymer solution.
This is because the density of segments
near a point which is free of
polymer is lower than in the bulk. The fractional
reduction in density near a
sphere/cavity is given by the term in parentheses on the right of
Eq. (\ref{y1}) \cite{eisenriegler96}.
It is illustrated in Fig. 2.
The methodology of the derivation of Eq. (\ref{y}) is quite general,
it may also be used to derive $y_P$ for ideal chains and for other
dimensionalities.

The positions
of the spheres in the polymer solution are correlated over a
range $\xi$; they are
distributed as if there is an attractive part to their interaction.
The effective attraction is expressed as an effective
potential between two spheres, called the potential of
mean force $w(r)$ \cite{hill60}. 
This is defined by
\begin{equation}
w(r)=-T\ln g_S(r)
\end{equation}
which is, using Eq. (\ref{y}),
\begin{equation}
w(r)\sim\left\{
\begin{array}{ll}
\infty & ~~~~~~ r<D\\
-T\ln\left[1+\frac{(D/\xi)^{8/3}}{\left(1-(D/\xi)^{4/3}\right)^2}
\left(g_P(r)-1\right)\right] &  ~~~~~~ r>D
\end{array}\right. .
\end{equation}
As $D/\xi\ll1$, we may linearise the logarithm
\begin{equation}
w(r)\sim-T\frac{(D/\xi)^{8/3}}{\left(1-(D/\xi)^{4/3}\right)^2}
\left(g_P(r)-1\right) ~~~~~~ r>D.
\label{wr}
\end{equation}
Then from Eqs. (\ref{gofrp}) and (\ref{wr}) we see that the depth of the
well in the potential of mean force is $T(D/\xi)^{4/3}\ll T$.
The potential is weak but long ranged, and as the range is increased the
depth of the well decreases.

Now that the distribution function between spheres has been obtained we can
calculate the structure factor $S(q)$, where $q$ is the wavevector.
$S(q)$ could be measured in a scattering
experiment performed on the mixture; note that it is not the total structure
factor for the mixture, just the partial structure factor of the spheres.
If we define the total correlation function
for the spheres $h_S(r)=g_S(r)-1$ \cite{hansen86}, then
\begin{equation}
S(q)=1+c_Sh_S(q).
\label{sofq}
\end{equation}
The Fourier transform of the total correlation function, $h_S(q)$, is
\begin{equation}
h_S(q)=\int h_S(r)e^{i{\bf q}.{\bf r}}d{\bf r},
\label{h1}
\end{equation}
where $h_P$ is the polymer
segment--segment total correlation function.
Using Eqs. (\ref{gy}) and (\ref{y}), Eq. (\ref{h1}) becomes
\begin{equation}
h_S(q)=-\int_0^D e^{i{\bf q}.{\bf r}} d{\bf r}
+\frac{(D/\xi)^{8/3}}{\left(1-(D/\xi)^{4/3}\right)^2}
\int_D^{\infty} h_P(r)e^{i{\bf q}.{\bf r}} d{\bf r},
\label{h2}
\end{equation}
where $h_P(r)=g_P(r)-1$.
The first integral of Eq. (\ref{h2}) is just the lowest
order, in density, approximation to the Fourier transform of
the total correlation function between hard spheres \cite{hansen86}.
The second term of Eq. (\ref{h2}) is not the Fourier transform of $h_P(r)$
because the region of integration is restricted to $r>D$. However,
this region only contributes a fraction $(D/\xi)^{5/3}$ to
the integrand, for $q\ll D^{-1}$, and so we approximate the 
integral of the second term
of Eq. (\ref{h2}) by $h_P(q)$. Then, Eq. (\ref{h2}) becomes
\begin{equation}
h_S(q)\sim-4\pi D^3\left(\frac{j_1(qD)}{qD}\right)
+\frac{(D/\xi)^{8/3}}{\left(1-(D/\xi)^{4/3}\right)^2}
h_P(q),
\label{hsq}
\end{equation}
where $j_1$ is the first
spherical Bessel function, $j_1(z)=\sin(z)/z^2-\cos(z)/z$.
We have related $h_S(q)$ and hence $S(q)$ to the polymer segment--segment
correlation function $h_P$ and so $h_P$, and therefore $\xi$,
can be measured
using only scattering from the spheres.

For $h_P(q)$ we use the Ornstein-Zernike form,
valid for small $q$, $q\ll\xi^{-1}$,
\begin{equation}
h_P(q)\sim \frac{\xi^3}{(q\xi)^2+1},
\label{hp}
\end{equation}
which is the Fourier transform of the third line of Eq. (\ref{gofrp}).
Although this gives the correct behaviour at small $q$, at $q\gg\xi$,
$h_P$ varies as $q^{-5/3}$ not the $q^{-2}$ given by Eq. (\ref{hp}):
$5/3$ is the exponent for a self-avoiding walk and at length scales
smaller than $\xi$ the segment--segment interactions are not screened
and the chains are self-avoiding \cite{degennes}.
We take Eq. (\ref{hp}) as unless the difference between $\xi$ and $D$ is
very large, say ${\cal O}(100)$ which requires a very large polymer,
then the behaviour at small enough values of $q$ to distinguish
between $5/3$ and 2 will be obscured by first term in Eq. (\ref{hsq})
and so not observable in measurements of $S(q)$.
Then,
\begin{equation}
h_S(q)\sim-4\pi D^3\left(\frac{j_1(qD)}{qD}\right)
+\frac{D^{8/3}\xi^{1/3}}{(1-(D/\xi)^{4/3})^2}
\left(\frac{1}{(q\xi)^2+1}\right).
\label{hofq}
\end{equation}

In Fig. 3 we have plotted $h_S(q)$ 
both with and without a semidilute
polymer solution.
The $h_S(q)$ is just what we would expect for particles interacting via
a short ranged repulsion --- the hard sphere interaction ---
and a long ranged
attraction --- the attractive potential of mean force due to the polymer.
Fig. 3 may be compared with the Fig. 5 of Khalatur {\it et al.}
\cite{khalatur97} who observe the same effect.
Note that at large values of $q$ the polymer has no effect,
$h_S$ is dominated at large $q$ by the hard-sphere interactions. While
at low $q$ $h_S$ is less negative when polymer is present.
We can compare this with a van der Waals
fluid in which the attractive interaction is infinitely long ranged;
in that case
$h_S(q)=h_{HS}(q)$, for $h_{HS}$ the hard-sphere total correlation function,
except for $q=0$ where $h_S>h_{HS}$ by an amount
proportional to the extent of the attraction \cite{hansen86}.
Due to the long range of the attraction, $\xi$, as compared to the
repulsion, $D$, the spheres behave very much like a van der Waals fluid.
Indeed, they are an even better approximation to a van der Waals fluid than
are simple fluids such as the noble gases. The ratio of
the range of the attractive interaction to that of the repulsive interaction
is smaller for noble gases than for the spheres in the polymer solution.
However, this is only true at low densities of spheres. When the
density of spheres becomes comparable to the density of polymer
(rescaled) polymer segments the correlations between polymer segments
will be affected by the spheres and $\xi$ reduced.
The distribution of spheres in the mixture is much more uniform than that
of the polymer, although as we have shown there are significant correlations
in the positions of the sphere.
So, we estimate that even if the the density of spheres is comparable to that
of the polymer segments the spheres will not greatly affect the correlations.
The situation where the density of spheres is high is quite subtle and so
far beyond this work.

The interactions studied here between spheres in a polymer solution are
similar to those between spheres in a near-critical binary mixture
\cite{burkhardt95}. There too the spheres behave as if they attracted each
other, due to correlations (inhomogeneities) in the solvent in which they
are suspended. Both these interactions are analogous to the quantum-mechanical
Casimir interaction \cite{burkhardt95}.

\section{Conclusion}

The pair distribution function of small hard spheres in a semidilute
polymer has been obtained, Eq. (\ref{hofq}), and
it has been shown how to determine the polymer's correlation length
from the structure factor of the spheres.
The correlations in distribution function function are long ranged, i.e.,
the ratio of the range
of the polymer induced effective
attraction, $\xi$,
to the range of the repulsion, $D$, is much larger than the
ratio of the attractive to the repulsive forces in
simple fluids such as the noble gases.
Note that this effective attraction is in a mixture where all
interactions are purely repulsive excluded volume interactions.
As the range of the
attraction is just $\xi$, it is a function of the polymer concentration and
can be varied at will within the semidilute regime. Also the depth
of the attractive well between two spheres varies with $\xi$;
as the range increases the well depth decreases.
Finally, we note that as
only the ratio in size between the particle and the polymer's
correlation length is important, the behaviour described above would
also occur if the polymer is not a flexible polymer but
a semiflexible biopolymer such as DNA or wormlike micelles
\cite{cates90,sear96} and the particles are much
larger particles with diameters of a few hundred nm.

\section*{Acknowledgments}

It is a pleasure to acknowledge D. Frenkel for pointing out to me
that the long range correlations
in a polymer solution induce interactions between particles embedded in it,
and J. Polson for
a careful reading of the manuscript.
I would like to thank The Royal Society for the award of a fellowship and
the FOM institute AMOLF for its hospitality.
The work of the FOM Institute is part of the research program of FOM
and is made possible by financial support from the
Netherlands Organisation for Scientific Research (NWO).

\newpage
\section*{Figure Captions}
\vspace*{0.2in}

\noindent
{\bf Fig. 1}
A schematic picture of a small colloidal sphere, a micelle
or a protein molecule, and a nearby part of a polymer chain.
The black circle is a spherical particle.
The chain is shown as a black curve and the dashed circles represent
the rescaled segments of size $D$.

\vspace*{0.1in}
\noindent
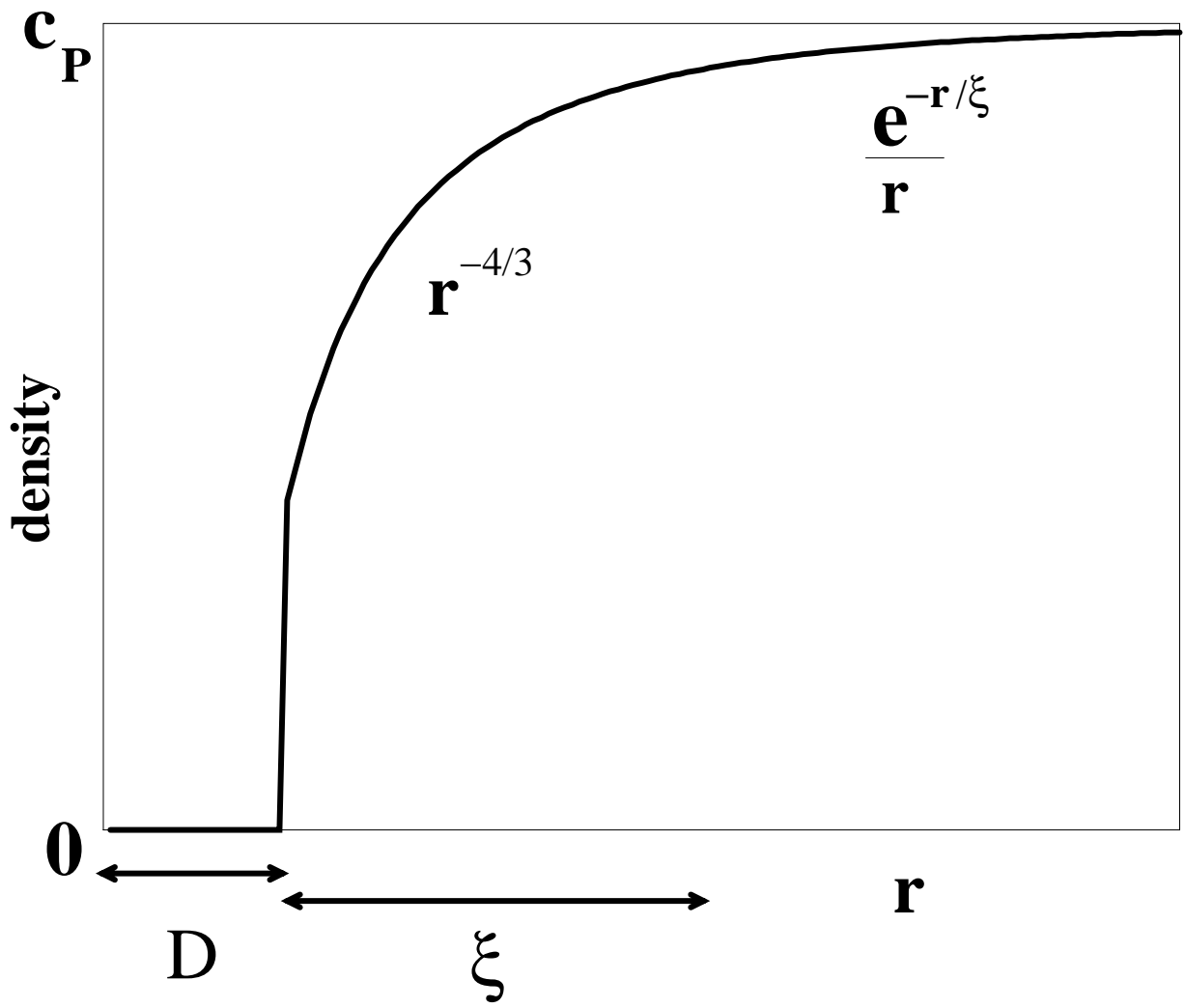
{\bf Fig. 2}
The polymer density around a sphere/cavity in the polymer solution.
If there is a sphere at the origin then there can be no polymer segment
there and so the polymer density is reduced in the surrounding volume.
The $x$-axis of the graph is $r$ the distance from the sphere and the
$y$-axis is the polymer density.
The segment density in contact with the sphere is less than $c_P$ but
non-zero; it relaxes to the bulk value $c_P$ at a rate $\sim r^{-4/3}$
for $r<\xi$ and exponentially when $r>\xi$.

\vspace*{0.1in}
\noindent
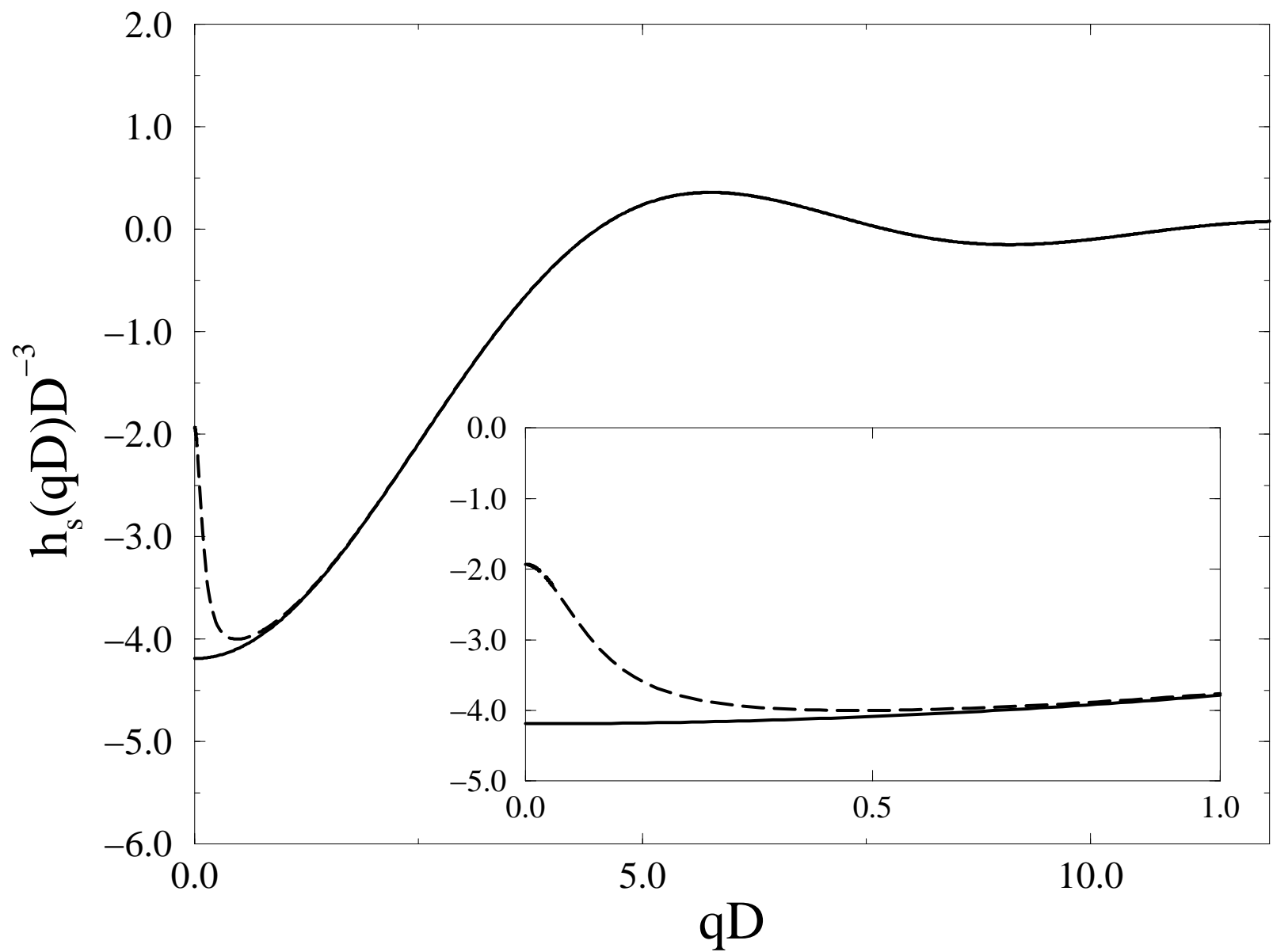
{\bf Fig. 3}
The total correlation function for low-density hard spheres.
The solid curve is for pure hard spheres and the dashed curve is for
spheres in a semidilute polymer solution with $\xi/D=10$.
The inset shows the small $q$ part at a larger scale.


\newpage

\begin{thebibliography}{99}

\bibitem{hill60} Hill, T. L.,
{\it Introduction to Statistical Thermodynamics}
(Addison-Wesley, Reading, Massachusetts, 1960).

\bibitem{degennes} de Gennes P.-G.,
{\it Scaling Concepts in Polymer Physics}
(Cornell, Ithaca, 1979).

\bibitem{cannell87} Cannell D. S., Wiltzius P. and Schaefer D. W.,
{\it Physics of Complex and Supermolecular Fluids},
S. A. Safran and N . A. Clark eds.
(John Wiley, New York, 1987).

\bibitem{abbott91} Abbott N. L., Blankschtein D. and Hatton T. A.,
{\it Macromolecules} {\bf 24} (1991) 4334.

\bibitem{abbott92} Abbott N. L., Blankschtein D. and Hatton T. A.,
{\it Macromolecules} {\bf 25} (1992) 5192.

\bibitem{clegg94} Clegg S. M., Williams P. A., Warren P. and Robb I. D.,
{\it Langmuir} {\bf 10} (1994) 3390.

\bibitem{robb95} Robb I. D., Williams P. A., Warren P. and Tanaka R.,
{\it J. Chem. Soc. Faraday Trans.} {\bf 91} (1995) 3901.

\bibitem{pusey91} Pusey P. N.,
{\it Liquids, Freezing and the Glass Transition},
J.-P. Hansen, D. Levesque and J. Zinn-Justin eds.
(North Holland, Amsterdam, 1991).

\bibitem{lekkerkerker95} Lekkerkerker H. N. W., Buining P., Buitenhuis J.,
Vroege G. J. and Stroobants A.,
{\it Observation, Prediction and Simulation of Phase Transitions in
Complex Fluids}, M. Baus {\it et al.} eds. (Kluwer, Dordrecht, 1995).

\bibitem{vrij76} Vrij A.,
{\it Pure Appl. Chem.} {\bf 48} (1976) 471.

\bibitem{yethiraj92} Yethiraj A., Hall C. K. and Dickman R.,
{\it J. Colloid Interface Sci.} {\bf 151} (1992) 102.

\bibitem{meijer94} Meijer E. J. and Frenkel D.,
{\it J. Chem. Phys.} {\bf 100} (1994) 6873.

\bibitem{khalatur97} Khalatur P. G., Zherenkova L. V. and Khokhlov A. R.,
{\it J. Phys. (France)} {\bf 7} (1997) 543.

\bibitem{degennes79} de Gennes P.-G.,
{\it C. R. Acad. Sci. Paris} {\bf 288B} (1979) 359.

\bibitem{odijk96} Odijk T.,
{\it Macromolecules} {\bf 29} (1996) 1842.

\bibitem{sear97} Sear R. P.,
{\it Phys. Rev. E} {\bf 56} (1997) 4463.

\bibitem{joanny79} Joanny J. F., Leibler L. and de Gennes P.-G.,
{\it J. Polymer Sci. Polymer Phys. Ed.} {\bf 17} (1979) 1073.

\bibitem{hansen86} Hansen J. P. and McDonald I. R.,
{\it Theory of Simple Liquids} (Academic Press, London, 2nd edn, 1986).

\bibitem{reiss92} Reiss H.,
{\it J. Phys. Chem.} {\bf 96} (1992) 4736.

\bibitem{eisenriegler96} Eisenriegler E., Hanke A. and Dietrich S.,
Phys. Rev. E {\bf 54} (1996) 1134.

\bibitem{burkhardt95} Burkhardt T. W. and Eisenriegler E.,
{\it Phys. Rev. Lett.} {\bf 74} (1995) 3189.

\bibitem{cates90} Cates M. E. and Candau S. J., {\it J. Phys. Condens. Matter}
{\bf 2} (1990) 6869.

\bibitem{sear96} Sear R. P. and Mulder B. M.,
{\it J. Phys. Chem. B} {\bf 101} (1997) 4839.

\end{thebibliography}
\end{document}